\documentclass[12pt]{article}
\usepackage{amssymb}
\renewcommand{\theequation}{\arabic{section}.\arabic{equation}}

\font\larl=cmr10 at 24pt

\newcommand{\vTb}{\vphantom{\mbox{\larl I}}}

\begin{document}



\def\a{\alpha}
\def\b{\beta}
\def\d{\delta}
\def\e{\epsilon}
\def\g{\gamma}
\def\h{\mathfrak{h}}
\def\k{\kappa}
\def\l{\lambda}
\def\o{\omega}
\def\p{\wp}
\def\r{\rho}
\def\t{\theta}
\def\s{\sigma}
\def\z{\zeta}
\def\x{\xi}
 \def\A{{\cal{A}}}
 \def\B{{\cal{B}}}
 \def\C{{\cal{C}}}
 \def\D{{\cal{D}}}
\def\G{\Gamma}
\def\K{{\cal{K}}}
\def\O{\Omega}
\def\R{\bar{R}}
\def\T{{\cal{T}}}
\def\L{\Lambda}
\def\f{E_{\tau,\eta}(sl_2)}
\def\E{E_{\tau,\eta}(sl_n)}
\def\Zb{\mathbb{Z}}
\def\Cb{\mathbb{C}}

\def\R{\overline{R}}

\def\beq{\begin{equation}}
\def\eeq{\end{equation}}
\def\bea{\begin{eqnarray}}
\def\eea{\end{eqnarray}}
\def\ba{\begin{array}}
\def\ea{\end{array}}
\def\no{\nonumber}
\def\le{\langle}
\def\re{\rangle}
\def\lt{\left}
\def\rt{\right}

\newtheorem{Theorem}{Theorem}
\newtheorem{Definition}{Definition}
\newtheorem{Proposition}{Proposition}
\newtheorem{Lemma}{Lemma}
\newtheorem{Corollary}{Corollary}
\newcommand{\proof}[1]{{\bf Proof. }
        #1\begin{flushright}$\Box$\end{flushright}}

\baselineskip=20pt

\newfont{\elevenmib}{cmmib10 scaled\magstep1}
\newcommand{\preprint}{
   \begin{flushleft}
   \end{flushleft}\vspace{-1.3cm}
   \begin{flushright}\normalsize
   \end{flushright}}
\newcommand{\Title}[1]{{\baselineskip=26pt
   \begin{center} \Large \bf #1 \\ \ \\ \end{center}}}
\newcommand{\Author}{\begin{center}
   \large \bf
Wen-Li Yang${}^{a,b}$~ and~Yao-Zhong Zhang ${}^b$\end{center}}
\newcommand{\Address}{\begin{center}

     ${}^a$ Institute of Modern Physics, Northwest University,
     Xian 710069, P.R. China\\
     ${}^b$ The University of Queensland, School of Mathematics and
        Physics, Brisbane, QLD 4072, Australia

   \end{center}}
\newcommand{\Accepted}[1]{\begin{center}
   {\large \sf #1}\\ \vspace{1mm}{\small \sf Accepted for Publication}
   \end{center}}

\preprint
\thispagestyle{empty}
\bigskip\bigskip\bigskip

\Title{Partition function of the eight-vertex model with domain
wall boundary condition} \Author

\Address
\vspace{1cm}

\begin{abstract}
We derive the recursive relations of the partition function for
the eight-vertex model on an $N\times N$ square lattice with
domain wall boundary condition. Solving the recursive relations,
we obtain the explicit expression of the domain wall partition
function of the model.  In the trigonometric/rational limit, our
results recover the corresponding ones for the six-vertex model.

\vspace{1truecm} \noindent {\it PACS:} 75.10.Pq, 04.20.Jb,
05.50.+q

\newpage
\end{abstract}
\newpage
\section{Introduction}
\label{intro} \setcounter{equation}{0}

The domain wall boundary condition (DW) for the six vertex model
on a finite square lattice was  introduced  by Korepin in
\cite{Kor82}, where some recursion relations of the partition
function which fully determine the partition function were also
derived. It was then found in \cite{Ize87,Ize92}  that the
partition function can be represented as a determinant. Such an
explicit expression of the partition function has played an
important role in constructing norms of Bethe states, correction
functions \cite{Ess95,Kor93,Kit99} and  thermodynamical properties
of the six-vertex model \cite{Kor00, Ble05}, and also in the  Toda
theories \cite{Sog93}. Moreover, it has been proven to be very
uesful in solving some pure mathematical problems, such as the
problem of alternating sign matrices \cite{Kup96}. Recently, the
partition functions with DW boundary condition have been obtained
for the high-spin models \cite{Car06} and the fermionic models
\cite{Zha07,Fod08}.

Among solvable models, elliptic ones stand out as a particularly
important class due to the fact that most trigonometric and
rational models can be obtained from them by  certain limits. In
this paper, we focus on the most fundamental elliptic model---the
eight-vertex model \cite{Bax71,Bax82} whose trigonometric limit
gives the six-vertex model. By means of the algebraic Bethe ansatz
method we derive an explicit expression of the partition function
for the eight-vertex model on an $N\times N$ square lattice with
the DW boundary condition. In the trigonometric limit, our results
recover those obtained by Korepin et al in
\cite{Kor82,Ize87,Ize92} for the six-vertex model.

The paper is organized as follows.  In section 2, we introduce our
notation and some basic ingredients. In section 3, after briefly
reviewing the vertex-face correspondence, we introduce the four
boundary states which specify the DW boundary condition of the
eight-vertex model. In section 4, some properties of the partition
function of the eight-vertex model with the DW boundary condition
are obtained by using algebraic Bethe ansatz method. With help of
these properties, we derive in section 5 the recursive relations
of the partition function and obtain the explicit expression of
the DW partition function by resolving the recursive relations. In
section 6, we summarize our results and give some discussions.
Some detailed technical proofs are given in Appendices A-B.


\section{The Eight-vertex model}
\label{XYZ} \setcounter{equation}{0}

In this section, we define the DW boundary condition for  the
eight-vertex model  on an $N\times N$ square lattice \cite{Bax82}.

\subsection{The eight-vertex R-matrix}

Let us fix $\tau$ such that ${\rm Im}(\tau)>0$ and a generic
complex number $\eta$. Introduce the following elliptic functions
\bea
\t\lt[\begin{array}{c} a\\b
  \end{array}\rt](u,\tau)&=&\sum_{n=-\infty}^{\infty}
  \exp\lt\{i\pi\lt[(n+a)^2\tau+2(n+a)(u+b)\rt]\rt\},\label{Function-a-b}\\
\theta^{(j)}(u)&=&\theta\lt[\begin{array}{c}\frac{1}{2}-\frac{j}{2}\\
 [2pt]\frac{1}{2}
 \end{array}\rt](u,2\tau),\quad j=1,2,\label{Function-j}\\[2pt]
\s(u)&=&\t\lt[\begin{array}{c}\frac{1}{2}\\[2pt]\frac{1}{2}
\end{array}\rt](u,\tau),\quad \s'(u)=\frac{\partial}{\partial u}
\lt\{\s(u)\rt\}.\label{Function} \eea The
$\s$-function\footnote{Our $\s$-function is the
$\vartheta$-function $\vartheta_1(u)$ \cite{Whi50}. It has the
following relation with the {\it Weierstrassian\/} $\s$-function
$\s_w(u)$: $\s_w(u)\propto e^{\eta_1u^2}\s(u)$ with
$\eta_1=\pi^2(\frac{1}{6}-4\sum_{n=1}^{\infty}\frac{nq^{2n}}{1-q^{2n}})
$ and $q=e^{i\tau}$.}
 satisfies the so-called Riemann
identity:\bea
&&\s(u+x)\s(u-x)\s(v+y)\s(v-y)-\s(u+y)\s(u-y)\s(v+x)\s(v-x)\no\\
&&~~~~~~=\s(u+v)\s(u-v)\s(x+y)\s(x-y),\label{identity}\eea which
will be useful in the following.

Let $V$ be a two-dimensional vector space $\Cb^2$ and
$\{\e_i|i=1,2\}$ be the orthonormal basis  of $V$ such that
$\langle \e_i,\e_j\rangle=\d_{ij}$. The well-known eight-vertex
model R-matrix $R(u)\in {\rm End}(V\otimes V)$ is given by \bea
R(u)=\lt(\begin{array}{llll}a(u)&&&d(u)\\&b(u)&c(u)&\\
&c(u)&b(u)&\\d(u)&&&a(u)\end{array}\rt). \label{r-matrix}\eea The
non-vanishing matrix elements  are \cite{Bax82}\bea
&&a(u)=\frac{\t^{(1)}(u)\,\t^{(0)}(u+\eta)\,\s(\eta)}
{\t^{(1)}(0)\, \t^{(0)}(\eta)\,\s(u+\eta)},\quad
b(u)=\frac{\t^{(0)}(u)\,
 \t^{(1)}(u+\eta)\,\s(\eta)}
{\t^{(1)}(0)\,\t^{(0)}(\eta)\,\s(u+\eta)},\no\\
&&c(u)=\frac{\t^{(1)}(u)\,
 \t^{(0)}(u+\eta)\,\s(\eta)}
{\t^{(1)}(0)\, \t^{(1)}(\eta)\,\s(u+\eta)},\quad
d(u)=\frac{\t^{(0)}(u)\,
 \t^{(0)}(u+\eta)\,\s(\eta)}
{\t^{(1)}(0)\t^{(1)}(\eta)\,\s(u+\eta)}.\label{r-func}\eea Here
$u$ is the spectral parameter and $\eta$ is the so-called crossing
parameter. The R-matrix satisfies the quantum Yang-Baxter equation
(QYBE) \bea R_{1,2}(u_1-u_2)R_{1,3}(u_1-u_3)R_{2,3}(u_2-u_3)
=R_{2,3}(u_2-u_3)R_{1,3}(u_1-u_3)R_{1,2}(u_1-u_2),\label{QYBE}\eea
and the properties, \bea &&\hspace{-1.5cm}\mbox{$Z_2$-symmetry}:
\, \qquad\s^i_1\s^i_2R_{1,2}(u)=R_{1,2}(u)\s^i_1\s^i_2,\quad
\mbox{for}\,\,
i=x,y,z,\label{Z2-sym}\\
 &&\hspace{-1.5cm}\mbox{Initial condition}:
 \,\,\,\,\,\,R_{1,2}(0)=P_{12},
\label{Initial} \eea Here $\s^{x},\s^{y},\s^{z}$ are the Pauli
matrices and  $P_{12}$ is the usual permutation operator.
Throughout this paper we adopt the standard notations: for any
matrix $A\in {\rm End}(V)$, $A_j$ is an embedding operator in the
tensor space $V\otimes V\otimes\cdots$, which acts as $A$ on the
$j$-th space and as identity on the other factor spaces;
$R_{i,j}(u)$ is an embedding operator of R-matrix in the tensor
space, which acts as identity on the factor spaces except for the
$i$-th and $j$-th ones.

\subsection{The model}
The partition function of a statistical model on a two-dimensional
lattice  is defined by the following:
\begin{eqnarray}
 Z=\sum \exp\{-\frac{E}{kT}\}, \nonumber
\end{eqnarray}
where $E$ is the energy of the system, $k$ is the Bolzmann
constant, $T$ is the temperature of the system,  and the summation
is taken over all possible configurations under the particular
boundary condition such as the DW boundary condition. The model we
consider here has eight allowed local vertex configurations

$$
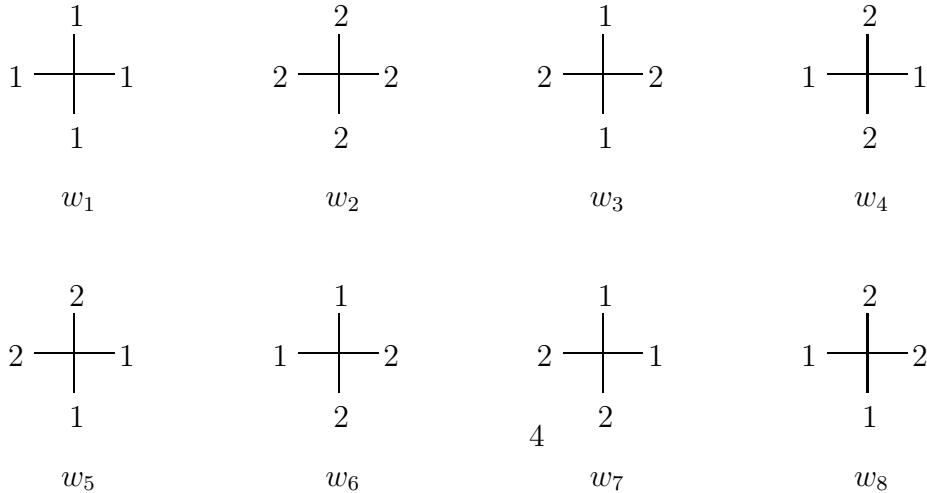
\begin{picture}(50,50)(0,0)
\put(-135,5){\line(-1,0){30}} \put(-150,20){\line(0,-1){30}}
\put(-152,-23){$1$}\put(-152,23){$1$}
\put(-175,0){$1$}\put(-133,0){$1$} \put(-155,-45){$w_1$}
\put(-35,5){\line(-1,0){30}} \put(-50,20){\line(0,-1){30}}
\put(-52,-23){$2$}\put(-52,23){$2$}
\put(-75,0){$2$}\put(-33,0){$2$} \put(-55,-45){$w_2$}
\put(65,5){\line(-1,0){30}} \put(50,20){\line(0,-1){30}}
\put(48,-23){$1$}\put(48,23){$1$} \put(25,0){$2$}\put(67,0){$2$}
\put(45,-45){$w_3$}
\put(165,5){\line(-1,0){30}} \put(150,20){\line(0,-1){30}}
\put(148,-23){$2$}\put(148,23){$2$}
\put(125,0){$1$}\put(167,0){$1$} \put(145,-45){$w_4$}
\end{picture}
$$\\[3mm]
$$
\begin{picture}(50,50)(0,0)
\put(-135,5){\line(-1,0){30}} \put(-150,20){\line(0,-1){30}}
\put(-152,-23){$1$}\put(-152,23){$2$}
\put(-175,0){$2$}\put(-133,0){$1$} \put(-155,-45){$w_5$}
\put(-35,5){\line(-1,0){30}} \put(-50,20){\line(0,-1){30}}
\put(-52,-23){$2$}\put(-52,23){$1$}
\put(-75,0){$1$}\put(-33,0){$2$} \put(-55,-45){$w_6$}
\put(65,5){\line(-1,0){30}} \put(50,20){\line(0,-1){30}}
\put(48,-23){$2$}\put(48,23){$1$} \put(25,0){$2$}\put(67,0){$1$}
\put(45,-45){$w_7$}
\put(165,5){\line(-1,0){30}} \put(150,20){\line(0,-1){30}}
\put(148,-23){$1$}\put(148,23){$2$}
\put(125,0){$1$}\put(167,0){$2$} \put(145,-45){$w_8$}
\end{picture}
$$\\[2mm]
$$
\begin{picture}(50,50)(0,0)
\put(-175,30){\small{\bf Figure 1.} Vertex configurations and
their associated Boltzmann weights.}
\end{picture}
$$

\noindent where $1$ and $2$ respectively denote the spin up and
down states. Each of these eight configurations is assigned a
statistical weight (or Boltzmann weight) $w_i$. Then the partition
function can be rewritten as
\begin{eqnarray}
Z=\sum\,{w_1}^{n_1}\,{w_2}^{n_2}\,{w_3}^{n_3}\,{w_4}^{n_4}\,{w_5}^{n_5}
 \,{w_6}^{n_6}\,{w_7}^{n_7}\,{w_8}^{n_8},\no
\end{eqnarray} where the summation is over all possible vertex configurations with $n_i$
being the number of the vertices of type $i$. If the local
Boltzmann weights have $Z_2$-symmetry, i.e.,
\begin{eqnarray}
  a\equiv w_1=w_2,\quad b\equiv w_3=w_4, \quad c\equiv
  w_5=w_6,\quad d\equiv w_7=w_8,
\end{eqnarray} and the variables $a,b,c,d$ satisfy a function
relation, or equivalently, the local Boltzmann weights $\{w_i\}$
can be parameterized by the matrix elements of the eight-vertex
R-matrix $R$ (\ref{r-matrix})-(\ref{r-func}) as in  figure 2,

$$
\begin{picture}(80,80)(0,0)
\put(-40,-15){\line(-1,0){60}} \put(-70,15){\line(0,-1){60}}
\put(-74,-60){$j$}\put(-74,20){$i$}
\put(-74,40){$\downarrow$}\put(-74,60){u}
\put(-110,-20){$l$}\put(-38,-20){$k$}
\put(-20,-20){$\leftarrow$}\put(0,-20){$\xi$} \put(32,-20){$=$}
\put(52,-20){$R^{j\,l}_{i\,k}(u-\xi)$,}
\put(120,-20){$i,j,k,l=1,2.$}
\end{picture}
$$ \vskip12mm
$$
\begin{picture}(50,50)(0,0)
\put(-175,20){\small{\bf Figure 2.} The Bolzmann weights and
elements of the eight-vertex R-matrix.}
\end{picture}
$$

\noindent then the corresponding model is called the eight-vertex
model which can be  exactly solved \cite{Bax82}. Therefore the
partition function of the eight-vertex model is give by
\begin{eqnarray}
Z=\sum\,{a}^{n_1+n_2}\,{b}^{n_3+n_4}\,{c}^{n_5+n_6}
 \,{d}^{n_7+n_8}.\no
\end{eqnarray}

In order to parameterize the local Boltzmann weights in terms of
the elements of the R-matrix, one needs to assign spectral
parameters $u$ and $\xi$ respectively to the vertical line and
horizontal line of each vertex of the lattice, as shown in  figure
2. In an inhomogeneous model, the statistical weights are
site-dependent. Hence two sets of spectral parameters
$\{u_{\alpha}\}$ and $\{\xi_i\}$ are needed, see figure 3. The
horizontal lines are enumerated by indices $1,\ldots,N$ with
spectral parameters $\{\xi_i\}$, while the vertical lines are
enumerated by indices $\bar{1},\ldots,\bar{N}$ with spectral
parameters $\{u_{\alpha}\}$. The DW boundary condition is
specified by four boundary states $|\Omega^{(1)}(\lambda)\rangle$,
$|\bar{\Omega}^{(1)}(\lambda)\rangle$,
$\langle\Omega^{(2)}(\lambda+\eta N\hat{2})|$ and
$\langle\bar{\Omega}^{(2)}(\lambda+\eta N\hat{2})|$ (the
definitions of the boundary states  will be given later in section
3, see (\ref{Boundary-state-1})-(\ref{Boundary-state-4}) below).
These four states correspond to the particular choices of spin
states on the four boundaries of the lattice. In contrast to the
six-vertex case \cite{Ess95}, our boundary states not only depend
on the spectral parameters ($|\Omega^{(1)}(\lambda)\rangle$ and
$\langle\Omega^{(2)}(\lambda+\eta N\hat{2})|$ depend on
$\{\xi_i\}$, while $|\bar{\Omega}^{(1)}(\lambda)\rangle$ and
$\langle\bar{\Omega}^{(2)}(\lambda+\eta N\hat{2})|$ depend on
$\{u_{\alpha}\}$) but also on two continuous parameters
$\lambda_1$ and $\lambda_2$ (it is  convenient to introduce a
vector $\lambda\in V$ associated with these two parameters
$\{\l_i\}$: $\l=\sum_{i=1}^2\l_i\e_i$). However, in the
trigonometric limit (i.e., setting $\l_2=\frac{\tau}{2}$ and then
taking $\tau\rightarrow +i\infty$), the corresponding boundary
states $|\Omega^{(1)}(\lambda)\rangle$ and
$\langle\bar{\Omega}^{(1)}( \lambda)|$ (or
$|\bar{\Omega}^{(2)}(\lambda+\eta N\hat{2})\rangle$ and
$\langle\Omega^{(2)}(\lambda+\eta N\hat{2})|$) become the state of
all spin up and its dual (or the state of all spin down and its
dual) up to some over-all scalar factors. Therefore the partition
function in the limit reduces to that of the six-vertex model
\cite{Kor82, Ize87,Ize92}. In this sense, we call the partition
function corresponding to the boundary condition given in figure 3
the DW partition function of the eight-vertex model.

\vskip20mm

\vskip3mm
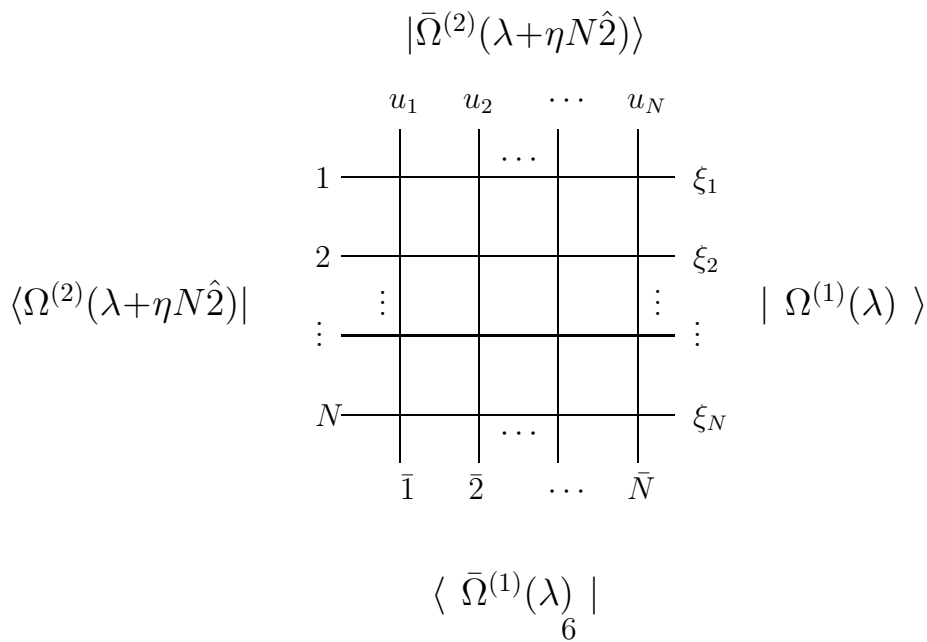
$$
\begin{picture}(160,160)(0,0)
\put(120,140){\line(-1,0){126}} \put(120,110){\line(-1,0){126}}
\put(120,80){\line(-1,0){126}} \put(120,50){\line(-1,0){126}}
 \put(106,158){\line(0,-1){126}}\put(76,158){\line(0,-1){126}}
 \put(46,158){\line(0,-1){126}}\put(16,158){\line(0,-1){126}}
 \put(127,136){$\xi_1$}\put(127,106){$\xi_2$}
 \put(112,88){$\vdots$}\put(8,88){$\vdots$}
 \put(152,88){{\large$|\hspace{+0.2truecm}
   \Omega^{(1)}(\lambda)\hspace{+0.2truecm}\rangle$}}
 \put(-132,88){{\large$\langle\Omega^{(2)}(\lambda
   \hspace{-0.1truecm}+\hspace{-0.1truecm}
   \eta N\hat{2})|$}}
 \put(127,76){$\vdots$}\put(127,46){$\xi_N$}
 \put(102,166){$u_N$}\put(72,166){$\cdots$}
 \put(54,144){$\cdots$}\put(54,40){$\cdots$}
 \put(40,166){$u_2$}\put(12,166){$u_1$}

 \put(102,18){$\bar{N}$}\put(72,18){$\cdots$}
 \put(42,18){$\bar{2}$}\put(16,18){$\bar{1}$}

 \put(-16,136){$1$}\put(-16,106){$2$}
 \put(-16,76){$\vdots$}\put(-16,46){$N$}

 \put(18,190){{\large$|\bar{\Omega}^{(2)}(\lambda
   \hspace{-0.1truecm}+\hspace{-0.1truecm}
   \eta N\hat{2})\rangle$}}
 \put(28,-22){{\large$\langle\hspace{+0.2truecm}\bar{\Omega}^{(1)}
   (\lambda)\hspace{+0.2truecm}|$}}
\end{picture}
$$ \vskip8mm
$$
\begin{picture}(50,50)(0,0)
\put(-175,40){\small{\bf Figure 3.} The eight-vertex model with DW
boundary condition.}
\end{picture}
$$


Now the partition function of the eight-vertex model with DW
boundary condition is a function of $2N+2$ variables $\{u_{\a}\}$,
$\{\xi_i\}$, $\l_1$ and $\l_2$, which will be denoted by
$Z_N(\{u_{\a}\};\{\xi_i\};\l)$. Due to the fact that the local
Boltzmann weights of each vertex of the lattice are given by the
matrix elements of the eight-vertex R-matrix (see figure 2), the
partition function can be expressed in terms of the product of the
R-matrices and the four boundary states \bea
 \hspace{-1truecm}Z_N(\{u_{\a}\};\{\xi_i\};\l)&=&\langle\Omega^{(2)}(\l+\eta
    N\hat{2})|\langle\bar{\Omega}^{(1)}(\l)|\,
    R_{\bar{1},N}(u_1-\xi_N)\ldots R_{\bar{1},1}(u_1-\xi_1)
    \ldots\no\\
 &&\,
    \times R_{\bar{N},N}(u_N-\xi_N)\ldots
    R_{\bar{N},1}(u_N-\xi_1)|\bar{\Omega}^{(2)}(\l+\eta N\hat{2})\rangle
    |\Omega^{(1)}(\l)\rangle.\label{PF}\eea
The aim of this paper is to obtain  an explicit expression for
$Z_N(\{u_{\a}\};\{\xi_i\};\l)$.

One can rearrange the product of the R-matrices in (\ref{PF}) in
terms of  a product  of the row-to-row monodromy matrices, namely,
\bea
 \hspace{-1truecm}Z_N(\{u_{\a}\};\{\xi_i\};\l)
 \hspace{-0.38truecm}&=&\hspace{-0.38truecm}\langle\Omega^{(2)}(\l
    \hspace{-0.1truecm}+\hspace{-0.1truecm}\eta
    N\hat{2})|\langle\bar{\Omega}^{(1)}(\l)|\,
    T_{\bar{1}}(u_1)\ldots T_{\bar{N}}(u_N)\,
    |\bar{\Omega}^{(2)}(\l
    \hspace{-0.1truecm}+\hspace{-0.1truecm}\eta N\hat{2})\rangle
    |\Omega^{(1)}(\l)\rangle,\label{PF-1}
\eea where the monodromy matrix $T_{\bar{i}}(u)$ is given by \bea
 T_{\bar{i}}(u)\equiv T_{\bar{i}}(u;\xi_1,\ldots,\xi_N)=R_{\bar{i},N}(u-\xi_N)\,\ldots
   R_{\bar{i},1}(u-\xi_1). \label{Monodromy-1}
\eea The QYBE (\ref{QYBE}) of the R-matrix gives rise to the
so-called ``RLL" relation satisfied by the monodromy matrix
$T_{\bar{i}}(u)$, \bea
 R_{\bar{i},\bar{j}}(u_i-u_j)\,T_{\bar{i}}(u_i)\,T_{\bar{j}}(u_j)
    =T_{\bar{j}}(u_j)\,T_{\bar{i}}(u_i)\,R_{\bar{i},\bar{j}}(u_i-u_j).\label{RLL-1}
\eea


\section{The boundary states}
\label{V-F} \setcounter{equation}{0}
\subsection{The Vertex-Face correspondence }
Let us briefly review the face-type R-matrix associated with the
eight-vertex model. From the orthonormal basis $\{\e_i\}$ of $V$,
we define \bea
 \hat{i}=\e_i-\overline{\e},~~\overline{\e}=
 \frac{1}{2}\sum_{k=1}^{2}\e_k,\qquad i=1,2,\qquad {\rm then}\quad
  \sum_{k=1}^2\hat{k}=0.
\eea For a generic $m\in V$, define \bea
 m_i=\langle m,\e_i\rangle, ~~m_{ij}=m_i-m_j=\langle
 m,\e_i-\e_j\rangle,~~i,j=1,2. \label{Def1}
\eea Let $\R(u;m)\in {\rm End}(V\otimes V)$ be the R-matrix of the
eight-vertex SOS model \cite{Bax82} given by \bea
 &&\R(u;m)\hspace{-0.1cm}=\hspace{-0.1cm}
 \sum_{i=1}^{2}\R^{ii}_{ii}(u;m)E_{ii}\hspace{-0.1cm}\otimes\hspace{-0.1cm}
 E_{ii}\hspace{-0.1cm}+\hspace{-0.1cm}\sum_{i\ne
 j}^2\lt\{\R^{ij}_{ij}(u;m)E_{ii}\hspace{-0.1cm}\otimes\hspace{-0.1cm}
 E_{jj}\hspace{-0.1cm}+\hspace{-0.1cm}
 \R^{ji}_{ij}(u;m)E_{ji}\hspace{-0.1cm}\otimes\hspace{-0.1cm}
 E_{ij}\rt\}, \label{R-matrix}
\eea where $E_{ij}$ is the matrix with elements
$(E_{ij})^l_k=\d_{jk}\d_{il}$. The coefficient functions are
\bea
 &&\R^{ii}_{ii}(u;m)=1,~~ \R^{ij}_{ij}(u;m)=\frac{\s(u)\s(m_{ij}-\eta)}
  {\s(u+\eta)\s(m_{ij})},~~i\neq j,\label{Elements1}\\
 && \R^{ji}_{ij}(u;m)=\frac{\s(\eta)\s(u+m_{ij})}
  {\s(u+\eta)\s(m_{ij})},~~i\neq j,\label{Elements2}
\eea  and $m_{ij}$ is defined in (\ref{Def1}). The R-matrix $\R$
satisfies the dynamical (modified) quantum Yang-Baxter equation
(or star-triangle equation) \cite{Bax82}.

Let us introduce two intertwiners which are $2$-component
column-vectors $\phi_{m,m-\eta\hat{j}}(u)$ labelled by
$\hat{\jmath}=\hat{1},\,\hat{2}$. The $k$-th element of
$\phi_{m,m-\eta\hat{j}}(u)$ is given by \bea
 \phi^{(k)}_{m,m-\eta\hat{j}}(u)=\theta^{(k)}(u+2m_j).\label{Int1}
\eea Explicitly,
\bea
 \phi_{m,m-\eta\hat{1}}(u)=
  \lt(\begin{array}{c}\t^{(1)}(u+2m_1)\\\\
  \t^{(2)}(u+2m_1)\end{array}\rt),\qquad
 \phi_{m,m-\eta\hat{2}}(u)=
  \lt(\begin{array}{c}\t^{(1)}(u+2m_2)\\\\
  \t^{(2)}(u+2m_2)\end{array}\rt).\no
\eea It is easy to check that these two intertwiner vectors
$\phi_{m,m-\eta\hat{i}}(u)$ are linearly {\it independent} for a
generic $m\in V$.

Using the intertwiner vectors, one can derive the following
vertex-face correspondence relation \cite{Bax82,Jim87} \bea
 &&R_{1,2}(u_1-u_2) \phi^{1}_{m,m-\eta\hat{\imath}}(u_1)\,
 \phi^2_{m-\eta\hat{\imath},m-\eta(\hat{\imath}+\hat{\jmath})}(u_2)
 \no\\&&~~~~~~= \sum_{k,l}\R(u_1-u_2;m)^{kl}_{ij}
 \phi^1_{m-\eta\hat{l},m-\eta(\hat{l}+\hat{k})}(u_1)\,
 \phi^2_{m,m-\eta\hat{l}}(u_2). \label{Face-vertex-1}
\eea Hereafter we adopt the convention: $\phi^1=\phi\otimes{\rm
id}\otimes \ldots$, $\phi^2={\rm id}\otimes \phi\otimes{\rm
id}\otimes \ldots$, etc. The QYBE (\ref{QYBE}) of the vertex-type
R-matrix $R(u)$ is equivalent to the dynamical Yang-Baxter
equation of the SOS R-matrix $\R(u;m)$. For a generic $m$, we can
introduce two row-vector intertwiners $\tilde{\phi}$ satisfying
the conditions, \bea
 && \tilde{\phi}_{m+\eta\hat{\mu},m}(u)
  ~\phi_{m+\eta\hat{\nu},m}(u)=\d_{\mu\nu},\quad \mu,\nu=1,2,
  \label{Int2}
\eea {}from which one  derives the relation,
\begin{eqnarray}
 &&\sum_{\mu=1}^2
 ~\phi_{m+\eta\hat{\mu},m}(u)\,\tilde{\phi}_{m+\eta\hat{\mu},m}(u)
 ={\rm id}.\label{Int4}\end{eqnarray}
With the help of (\ref{Int1})-(\ref{Int4}), we obtain the
following relations from the vertex-face correspondence relation
(\ref{Face-vertex-1}): \bea
&&\tilde{\phi}^1_{m+\eta\hat{k},m}(u_1)\,\,
 R_{1,2}(u_1-u_2)\,\, \phi^2_{m+\eta\hat{\jmath},m}(u_2)\no\\[6pt]
&&\qquad\quad= \sum_{i,l}\R(u_1-u_2;m)^{kl}_{ij}\,
 \tilde{\phi}^1_{m+\eta(\hat{\imath}+\hat{\jmath}),m+\eta\hat{\jmath}}(u_1)\,
 \phi^2_{m+\eta(\hat{k}+\hat{l}),m+\eta\hat{k}}(u_2),\label{Face-vertex-2}\\[6pt]
&&\tilde{\phi}^1_{m+\eta\hat{k},m}(u_1)\,
 \tilde{\phi}^2_{m+\eta(\hat{k}+\hat{l}),m+\eta\hat{k}}(u_2)\,\,R_{1,2}(u_1-u_2)\no\\[6pt]
&&\qquad\quad= \sum_{i,j}\R(u_1-u_2;m)^{kl}_{ij}\,
 \tilde{\phi}^1_{m+\eta(\hat{\imath}+\hat{\jmath}),m+\eta\hat{\jmath}}(u_1)
 \,\tilde{\phi}^2_{m+\eta\hat{\jmath},m}(u_2).\label{Face-vertex-3} \eea
The intertwiners $\phi$, $\tilde{\phi}$ and the associated
vertex-face correspondence relations will play an important role
in determining the very properties of the partition function
$Z_N(\{u_{\a}\};\{\xi_i\};\l)$ that enable us in section 4 to
fully determine its explicit expression.


\subsection{The boundary states}
Now we are in the position to construct the boundary states which
have been  used in section 2 to specify the DW boundary condition
of the eight-vertex model, see figure 3.

For any vector $m\in V$, we  introduce four states\footnote{Among
them, $|\Omega^{(i)}(m)\rangle$, with special choices of $m$, are
the complete reference states of the open XYZ spin chain
\cite{Yan06}, and have played an important role in constructing
the extra center elements of the elliptic algebra at roots of
unity \cite{Bel04,Yan05}}
 which live in the two $N$-tensor spaces of
$V$ (one is indexed by $1,\ldots,N$ and the other is indexed by
$\bar{1},\ldots,\bar{N}$) or their dual spaces as follows:
\bea
 |\Omega^{(i)}(m)\rangle&=&
   \phi^{1}_{m,m-\eta\hat{i}}(\xi_1)
   \phi^{2}_{m-\eta\hat{i},m-2\eta\hat{i}}(\xi_2)\ldots
   \phi^{N}_{m-\eta (N-1)\hat{i},m-\eta N\hat{i}}(\xi_N),\quad
   i=1,2,\label{Boundary-state-1}\\[8pt]
|\bar{\Omega}^{(i)}(m)\rangle&=&\phi^{\bar{1}}_{m,m-\eta\hat{i}}(u_1)
 \phi^{\bar{2}}_{m-\eta\hat{i},m-2\eta\hat{i}}(u_2) \ldots
 \phi^{\bar{N}}_{m-\eta (N-1)\hat{i},m-\eta N\hat{i}}(u_N),\quad
 i=1,2,\label{Boundary-state-2}\\[8pt]
\langle\Omega^{(i)}(m)|&=&
 \tilde{\phi}^{1}_{m,m-\eta\hat{i}}(\xi_1)
 \tilde{\phi}^{2}_{m-\eta\hat{i},m-2\eta\hat{i}}(\xi_2)\ldots
 \tilde{\phi}^{N}_{m-\eta(N-1)\hat{i},m-\eta N\hat{i}}(\xi_N),\quad
   i=1,2,\label{Boundary-state-3}\\[8pt]
\langle\bar{\Omega}^{(i)}(m)|&=&
 \tilde{\phi}^{\bar{1}}_{m,m-\eta\hat{i}}(u_1)
 \tilde{\phi}^{\bar{2}}_{m-\eta\hat{i},m-2\eta\hat{i}}(u_2)\ldots
 \tilde{\phi}^{\bar{N}}_{m-\eta(N-1)\hat{i},m-\eta N\hat{i}}(u_N),\quad
   i=1,2.\label{Boundary-state-4}
\eea The boundary states which have been used to define the DW
boundary condition in section 2 can be obtained through the above
states by special choices of $m$ and $i$ (for example,  $m$ is
specified to $\l$ or $\l+\eta N\hat{2}$). Then the DW partition
function $Z_N(\{u_{\a}\};\{\xi_i\};\l)$ given by (\ref{PF})
becomes \bea
 \hspace{-0.2truecm}Z_N(\{u_{\a}\};\{\xi_i\};\l)
 \hspace{-0.32truecm}&=&\hspace{-0.32truecm}
  \tilde{\phi}^1_{\l+\eta N\hat{2},\l+\eta
  (N-1)\hat{2}}(\xi_1)\ldots \tilde{\phi}^N_{\l+\eta \hat{2},\l}(\xi_N)
  \,\,\tilde{\phi}^{\bar{1}}_{\l,\l-\eta
  \hat{1}}(u_1)\ldots \tilde{\phi}^{\bar{N}}
  _{\l-\eta(N-1)\hat{1},\l-\eta N\hat{1}}(u_N)\no\\
 &&\times R_{\bar{1},N}(u_1-\xi_N)\ldots R_{\bar{1},1}(u_1-\xi_1)
    \ldots\,\,
    R_{\bar{N},N}(u_N-\xi_N)\ldots
    R_{\bar{N},1}(u_N-\xi_1)\no\\
 &&\times\phi^{\bar{1}}_{\l+\eta N\hat{2},\l+\eta(N-1)\hat{2}}(u_1)
    \ldots\phi^{\bar{N}}_{\l+\eta\hat{2},\l}(u_N)\,\,
    \phi^{1}_{\l,\l-\eta\hat{1}}(\xi_1)\ldots
    \phi^{N}_{\l-\eta (N-1)\hat{1},\l-\eta N\hat{1}}(\xi_N).\no\\
 && \label{PF-2}
\eea


\section{The properties of the partition function}
\label{Properties} \setcounter{equation}{0}

In this section we will derive certain properties of the DW
partition function $Z_N(\{u_{\a}\};\{\xi_i\};\l)$ which enable us
to determine its explicit expression.

For the case of $N=1$, the corresponding partition function
(\ref{PF-2}) becomes
\bea
 Z_1(u_1;\xi_1;\l)&=&\tilde{\phi}^1_{\l+\eta\hat{2},\l}(\xi_1)\,
   \tilde{\phi}^{\bar{1}}_{\l,\l-\eta\hat{1}}(u_1)
   \,R_{\bar{1},1}(u_1-\xi_1)\,\phi^{\bar{1}}_{\l+\eta\hat{2},\l}(u_1)
   \,\phi^1_{\l,\l-\eta\hat{1}}(\xi_1)\no\\
 &\stackrel{(\ref{Face-vertex-1})}{=}&\sum_{k,l=1}^2
  \,\R_{2\,1}^{k\,l}(u_1-\xi_1;\l+\eta\hat{2})
  \lt(\tilde{\phi}^{1}_{\l+\eta\hat{2},\l}(\xi_1)
  \phi^1_{\l+\eta\hat{2},\l+\eta(\hat{2}-\hat{l})}(\xi_1)\rt)\no\\
  &&\qquad \times\lt(\tilde{\phi}^{\bar{1}}_{\l,\l-\eta\hat{1}}(u_1)
  \phi^{\bar{1}}_{\l+\eta(\hat{k}-\hat{1}),\l-\eta\hat{1}}(u_1)\rt)
  \no\\
 &\stackrel{(\ref{Int2})}{=}&
 \R_{2\,1}^{1\,2}(u_1-\xi_1;\l+\eta\hat{2}).\no
\eea Thus, we have the first property of the partition function:
\bea
 Z_1(u_1;\xi_1;\l)=\frac{\s(\eta)\,\s(u_1-\xi_1+\l_{21}+\eta)}
  {\s(u_1-\xi_1+\eta)\,\s(\l_{21}+\eta)},\label{Property-1}
\eea where $\l_{21}$ is defined by (\ref{Def1}). Using the
fundamental exchange relation (\ref{RLL-1}) and the vertex-face
relations (\ref{Face-vertex-1}), (\ref{Face-vertex-2}) and
(\ref{Face-vertex-3}), we derive the second property of the
partition function: \bea
 &&\mbox{$Z_N(\{u_{\a}\};\{\xi_i\};\l)$ is a symmetric function of
 $\{u_{\a}\}$ and  $\{\xi_i\}$ separatively}.\label{Symmetry}
 \eea The proof of the
 above property is relegated to Appendix A.

 In addition to the Riemann identity (\ref{identity}), the
$\s$-function enjoys the following quasi-periodic properties:
\bea
 \s(u+1)&=&-\s(u),\quad
 \s(u+\tau)=-e^{-2i\pi(u+\frac{\tau}{2})}\s(u),
 \label{quasi-func}
\eea which are useful in deriving the quasi-periodicity of the
partition function. The expansions of the boundary states in terms
of the intertwiner vectors
(\ref{Boundary-state-1})-(\ref{Boundary-state-4}) and the
partition function in terms of the monodromy matrices (\ref{PF-1})
allow us to rewrite $Z_N(\{u_{\a}\};\{\xi_i\};\l)$ as \bea
 \hspace{-0.2truecm}Z_N(\{u_{\a}\};\{\xi_i\};\l)
 \hspace{-0.32truecm}&=&\hspace{-0.32truecm}
  \tilde{\phi}^1_{\l+\eta N\hat{2},\l+\eta(N-1)\hat{2}}(\xi_1)\ldots
  \tilde{\phi}^N_{\l+\eta \hat{2},\l}(\xi_N)\,
  \tilde{\phi}^{\bar{1}}_{\l,\l-\eta\hat{1}}(u_1)T_{\bar{1}}(u_1)
  \phi^{\bar{1}}_{\l+\eta N\hat{2},\l+\eta(N-1)\hat{2}}(u_1)\ldots\no\\
&&\hspace{-0.2truecm}\times\tilde{\phi}^{\bar{N}}_{\l-\eta(N-1)\hat{1},\l-\eta
   N\hat{1}}(u_N)T_{\bar{N}}(u_N)
  \phi^{\bar{N}}_{\l+\eta\hat{2},\l}(u_N)
  \phi^{1}_{\l,\l-\eta\hat{1}}(\xi_1)\ldots
    \phi^{N}_{\l-\eta (N-1)\hat{1},\l-\eta N\hat{1}}(\xi_N).\no
\eea The dependence of $Z_N(\{u_{\a}\};\{\xi_i\};\l)$ on the
argument $u_N$ only comes from the last term corresponding to the
second line of the above equation. Let us denote the term by
$A(u_N)$, namely, \bea
 A(u_N)&=&\tilde{\phi}^{\bar{N}}_{\l-\eta(N-1)\hat{1},\l-\eta
   N\hat{1}}(u_N)T_{\bar{N}}(u_N)
  \phi^{\bar{N}}_{\l+\eta\hat{2},\l}(u_N)
  \phi^{1}_{\l,\l-\eta\hat{1}}(\xi_1)\ldots
    \phi^{N}_{\l-\eta (N-1)\hat{1},\l-\eta N\hat{1}}(\xi_N)\no\\
  &=&\tilde{\phi}^{\bar{N}}_{\l-\eta(N-1)\hat{1},\l-\eta
   N\hat{1}}(u_N)R_{\bar{N},N}(u_N-\xi_N)
   \phi^{N}_{\l-\eta (N-1)\hat{1},\l-\eta
   N\hat{1}}(\xi_N)\ldots\no\\
 &\stackrel{(\ref{Face-vertex-2})}{=}&
    \phi^N_{\l-\eta(N-2)\hat{1},\l-\eta(N-1)\hat{1}}(\xi_N)
   \tilde{\phi}^{\bar{N}}_{\l-\eta(N-2)\hat{1},\l-\eta(N-1)\hat{1}}(u_N)
   R_{\bar{N},N-1}(u_N-\xi_{N-1})\ldots\no\\
   &&+\R(u_N-\xi_N;\l-\eta N\hat{1})^{12}_{21}
   \phi^N_{\l-\eta(N-2)\hat{1}+\eta\hat{2},\l-\eta(N-1)\hat{1}}(\xi_N)
   \tilde{\phi}^{\bar{N}}_{\l-\eta(N-2)\hat{1}+\eta\hat{2},\l-\eta(N-1)\hat{1}}(u_N)
   \ldots\no\\
   &=&\no\\
   &\vdots&\no\\
 &=&\phi^N_{\l-\eta(N-2)\hat{1},\l-\eta(N-1)\hat{1}}(\xi_N)
   \tilde{\phi}^{\bar{N}}_{\l-\eta(N-2)\hat{1},\l-\eta(N-1)\hat{1}}(u_N)
   R_{\bar{N},N-1}(u_N-\xi_{N-1})\ldots\no\\
 &&+\R(u_N-\xi_N;\l-\eta N\hat{1})^{12}_{21}\prod_{l=1}^{N-1}
    \R(u_N-\xi_l;\l-\eta l\hat{1})^{21}_{21}
    \phi^1_{\l+\eta\hat{2},\l+\eta\hat{2}-\eta\hat{1}}(\xi_1)\ldots\no
 \eea It can be shown by induction that  $A(u_N)$ satisfies the
 following quasi-periodicity:
\bea
  A(u_N+1)=A(u_N),\quad
  A(u_N+\tau)=e^{-2i\pi(\l_{21})}\,A(u_N).\no
\eea Since $Z_N(\{u_{\a}\};\{\xi_i\};\l)$ is a symmetric function
of $\{u_{\a}\}$, we conclude that $Z_N(\{u_{\a}\};\{\xi_i\};\l)$
has the following quasi-periodic properties
\bea
 &&\hspace{-1.2truecm}Z_N(u_1,\ldots,u_l+1,u_{l+1},\ldots;\{\xi_i\};\l)=Z_N(\{u_{\a}\};\{\xi_i\};\l),
    \quad\quad l=1,\ldots,N,\label{quasi-per-1}\\
 &&\hspace{-1.2truecm}Z_N(u_1,\ldots,u_l+\tau,u_{l+1},\ldots;\{\xi_i\};\l)
   =e^{-2i\pi(\l_{21})}\,Z_N(\{u_{\a}\};\{\xi_i\};\l),
   \, l=1,\ldots,N.\label{quasi-per-2}
\eea Using the expressions (\ref{Elements1})-(\ref{Elements2}) of
the matrix elements of the R-matrix $\R$ and the vertex-face
correspondences (\ref{Face-vertex-1}) and (\ref{Face-vertex-2}),
we find the analytic property of the partition function: \bea
  &&\mbox{$Z_N(\{u_{\a}\};\{\xi_i\};\l)$ is an analytic function of
    $u_l$ with simple poles $\{\xi_i-\eta|i=1,\ldots,N\}$ }\no\\
  &&\qquad\mbox{inside the fundamental (upright)
     rectangle \cite{Bax82} generated by $1$ and
     $\tau$}.\label{Analytic}
\eea
Direct calculation (for details see Appendix B) shows that at
each simple pole $\xi_i-\eta$ the corresponding residue is \bea
 {\rm Res}_{u_l=\xi_i-\eta}\lt(Z_N(\{u_{\a}\};\{\xi_j\};\l)\rt)&=&
   \frac{\s(\eta)\s(\l_{21})}{\s'(0)\s(\l_{21}+\eta)}
   \prod_{\a\neq l}\frac{\s(u_{\a}-\xi_i)}{\s(u_{\a}-\xi_i+\eta)}
   \prod_{j\neq i}\frac{\s(\xi_j-\xi_i+\eta)}{\s(\xi_j-\xi_i)}\no\\
 &&\times Z_{N-1}(\{u_{\a}\}_{\a\neq l};\{\xi_j\}_{j\neq
   i};\l+\eta\hat{2}),\, l,i=1,\ldots,N.\label{Residue}
\eea Similarly, using the initial condition (\ref{Initial}) of the
R-matrix $R$ and the vertex-face correspondences
(\ref{Face-vertex-1}) and (\ref{Face-vertex-2}), one can also show
that  the partition function $Z_N(\{u_{\a}\};\{\xi_i\};\l)$
satisfies  the following relations \bea
 Z_N(\{u_{\a}\};\{\xi_j\};\l)\lt|_{u_l=\xi_i}\rt.=
  Z_{N-1}(\{u_{\a}\}_{\a\neq l};\{\xi_j\}_{j\neq i};\l),\quad
  l,i=1,\ldots,N.\label{property-2}
\eea

 So remarks are in order. The properties (\ref{Property-1}),
 (\ref{Symmetry}), (\ref{quasi-per-1}), (\ref{quasi-per-2}),
 (\ref{Analytic}) and (\ref{Residue}) uniquely determine the
 partition function. On the other hand, the properties (\ref{Property-1}),
 (\ref{Symmetry}), (\ref{quasi-per-1}), (\ref{quasi-per-2}),
 (\ref{Analytic}) and (\ref{property-2}) also fully fix the
 partition function. They yield the recursive relations (\ref{recursive-1})
 and (\ref{recursive-2}) (see below) respectively.


\section{The partition function}
\label{Part-F} \setcounter{equation}{0}

In this section, we will derive two recursive relations from the
properties of the partition function obtained in the previous
section. Each of the recursive relations together with
(\ref{Property-1}) uniquely determines the partition function.

\subsection{The recursive relation}

We now concentrate on the $u_N$-dependence of  the partition
function $Z_N(\{u_{\a}\};\{\xi_j\};\l)$. {}From
(\ref{quasi-per-1}) and (\ref{quasi-per-2}), we have
\bea
 &&Z_N(u_1,\ldots,u_{N-1},u_N+1;\{\xi_j\};\l)=Z_N(\{u_{\a}\};\{\xi_j\};\l),\no\\
 &&Z_N(u_1,\ldots,u_{N-1},u_N+\tau;\{\xi_j\};\l)=e^{-2i\pi(\l_{21})}
     Z_N(\{u_{\a}\};\{\xi_j\};\l).\label{quasi-per-3}
\eea The analytic properties (\ref{Analytic}) and (\ref{Residue})
imply that \bea
 Z_N(\{u_{\a}\};\{\xi_j\};\l)&=&\sum_{i=1}^N\lt\{
  \frac{\s(\eta)\s(u_N-\xi_i+a_i)}{\s(u_N-\xi_i+\eta)\s(b_i)}
  \prod_{j\neq i}\frac{\s(\xi_j-\xi_i+\eta)}{\s(\xi_j-\xi_i)}
  \prod_{l\neq N}\frac{\s(u_l-\xi_i)}{\s(u_l-\xi_i+\eta)}\rt.\no\\
 && \times\lt.Z_{N-1}(\{u_{\a}\}_{\a\neq N};\{\xi_j\}_{j\neq
  i};\l+\eta\hat{2})\vTb\rt\}+\Delta,\no
\eea where $\{a_i\}$, $\{b_i\}$ and $\Delta$ are some constants
with respect to $u_N$, and $a_i$ and $b_i$ satisfy the constraints
\bea
 \frac{\s(a_i-\eta)}{\s(b_i)}=\frac{\s(\l_{21})}{\s(\l_{21}+\eta)}, \quad
 i=1,\ldots,N.\label{constraint-1}
\eea The quasi-periodic condition (\ref{quasi-per-3}) leads to
\bea
 \lt\{\begin{array}{ll}a_i=\l_{21}+\eta&i=1,\ldots,N,\\
       \Delta=0.&\end{array}\rt.
\eea The constraint (\ref{constraint-1}) then  yields that
$b_i=a_i=\l_{21}+\eta$. Thus the partition function
$Z_N(\{u_{\a}\};\{\xi_j\};\l)$ satisfies the following recursive
relation \bea
 Z_N(\{u_{\a}\};\{\xi_j\};\l)&=&\sum_{i=1}^N\lt\{
  \frac{\s(\eta)\s(u_N-\xi_i+\l_{21}+\eta)}{\s(u_N-\xi_i+\eta)\s(\l_{21}+\eta)}
  \prod_{j\neq i}\frac{\s(\xi_j-\xi_i+\eta)}{\s(\xi_j-\xi_i)}
  \prod_{l\neq N}\frac{\s(u_l-\xi_i)}{\s(u_l-\xi_i+\eta)}\rt.\no\\
 &&\quad\quad \times\lt.Z_{N-1}(\{u_{\a}\}_{\a\neq N};\{\xi_j\}_{j\neq
  i};\l+\eta\hat{2})\vTb\rt\}.\label{recursive-1}
\eea

On the other hand, the quasi-periodicity (\ref{quasi-per-3}) of
the partition function, the fact that the partition function only
has simple poles at $\{\xi_i-\eta\}$ and the relation
(\ref{property-2}) imply that the partition function has the
following expansion \bea
 Z_N(\{u_{\a}\};\{\xi_j\};\l)&=&\sum_{i=1}^N\lt\{
  \frac{\s(\eta)\s(u_N-\xi_i+a'_i)}{\s(u_N-\xi_i+\eta)\s(a'_i)}
  \prod_{j\neq i} \frac{\s(u_N-\xi_j)\s(\xi_i-\xi_j+\eta)}
  {\s(u_N-\xi_j+\eta)\s(\xi_i-\xi_j)}\,
  \rt.\no\\
 &&\quad\quad \times\lt.Z_{N-1}(\{u_{\a}\}_{\a\neq N};\{\xi_j\}_{j\neq
  i};\l)\vTb\rt\},\no
\eea where $\{a_i\}$ are some constants with respect to $u_N$. The
quasi-periodicity (\ref{quasi-per-3}) further requires
$a'_i=\l_{21}+N\eta$. Namely, the partition function
$Z_N(\{u_{\a}\};\{\xi_j\};\l)$ satisfies the following recursive
relation \bea
 Z_N(\{u_{\a}\};\{\xi_j\};\l)&=&\sum_{i=1}^N\lt\{
  \frac{\s(\eta)\s(u_N-\xi_i+\l_{21}+N\eta)}{\s(u_N-\xi_i+\eta)\s(\l_{21}+N\eta)}
  \prod_{j\neq i} \frac{\s(u_N-\xi_j)\s(\xi_i-\xi_j+\eta)}
  {\s(u_N-\xi_j+\eta)\s(\xi_i-\xi_j)}
  \rt.\no\\
 &&\quad\quad \times\lt.Z_{N-1}(\{u_{\a}\}_{\a\neq N};\{\xi_j\}_{j\neq
  i};\l)\vTb\rt\}.\label{recursive-2}
\eea In the trigonometric limit, the recursive relation
(\ref{recursive-2}) recovers that of \cite{Kit99} for the
six-vertex model.

\subsection{The DW partition function}

The recursive relation (\ref{recursive-1}) and the property
(\ref{Property-1}) uniquely determine
$Z_N(\{u_{\a}\};\{\xi_j\};\l)$, on the other hand the recursive
relation (\ref{recursive-2}) and the property (\ref{Property-1})
also fully fix the partition function. Using the Riemann identity
(\ref{identity}) of the $\s$-function, one can check that the
solution to each of recursive relations (\ref{recursive-1}) and
(\ref{recursive-2}) gives rise to a symmetric function of
$\{u_{\a}\}$ as required. As a consequence,  the two expressions
of the partition function obtained by solving the recursive
relations (\ref{recursive-1}) and (\ref{recursive-2}) respectively
are equal since they are related to each other by some permutation
of $\{u_{\a}\}$. Here, we present the result by resolving the
recursive relation (\ref{recursive-2}).

Using the property (\ref{Property-1}) and the recursive relation
(\ref{recursive-2}), we obtain the explicit expression of the
partition function $Z_N(\{u_{\a}\};\{\xi_j\};\l)$
\bea
 Z_N(\{u_{\a}\};\{\xi_j\};\l)&=&\sum_{s\in S_N}\prod_{l=1}^N
  \lt\{\frac{\s(\eta)\s(u_l-\xi_{s(l)}+\l_{21}+l\eta)}
    {\s(u_l-\xi_{s(l)}+\eta)\s(\l_{21}+l\eta)}\rt.\no\\
    &&\qquad\qquad\times \prod_{k=1}^{l-1}\lt.
    \frac{\s(u_l-\xi_{s(k)})\s(\xi_{s(l)}-\xi_{s(k)}+\eta)}
    {\s(u_l-\xi_{s(k)}+\eta)\s(\xi_{s(l)}-\xi_{s(k)})}\rt\},
    \label{partition-function}
\eea where $S_N$ is the permutation group of $N$ indices. It is
easy to see from the above explicit expression that
$Z_N(\{u_{\a}\};\{\xi_j\};\l)$ is indeed a symmetry function of
$\{\xi_i\}$.


\section{Conclusions and discussions}
\label{Con} \setcounter{equation}{0}

We have introduced the DW boundary condition specified by the four
boundary states (\ref{Boundary-state-1})-(\ref{Boundary-state-4})
for the eight-vertex model on an $N\times N$ square lattice. The
boundary states are the two-parameter generalization of the
all-spin-up and all-spin-down states and their dual states. With
the DW boundary condition, we have obtained the properties
(\ref{Property-1}), (\ref{Symmetry}), (\ref{quasi-per-1}),
(\ref{quasi-per-2}), (\ref{Residue}) and (\ref{property-2}) of the
partition function $Z_N(\{u_{\a}\};\{\xi_j\};\l)$. These
properties enable us to derive the two recursive relations
(\ref{recursive-1}) and (\ref{recursive-2}) of
$Z_N(\{u_{\a}\};\{\xi_j\};\l)$, whose trigonometric limits recover
those corresponding to the six-vertex model. The recursive
relation (\ref{recursive-1}) or (\ref{recursive-2}) together with
(\ref{Property-1}) uniquely determines the  partition function
$Z_N(\{u_{\a}\};\{\xi_j\};\l)$. Solving the recursive relations,
we obtain the explicit expression (\ref{partition-function}) of
the partition function.

Since the DW partition function (\ref{partition-function}) of the
eight-vertex model  reduces in the trigonometric limit to that of
the six-vertex model which can be expressed in terms of a
determinant \cite{Ize92}, it is natural to think that
(\ref{partition-function}) might be expressed in terms of a
determinant. However, it seems that the partition function
(\ref{partition-function}) could only be expressed in terms of a
sum of $N$ determinants as follows: \bea
 Z_N(\{u_{\a}\};\{\xi_j\};\l)&=&\frac{\prod_{\a=1}^N\prod_{j=1}^N\s(u_{\a}-\xi_j)}
    {\prod_{\a>\b}\s(u_{\a}-u_{\b})\prod_{k<l}\s(\xi_k-\xi_l)}
    \frac{1}{\s(\sum_{\a}(u_{\a}-\xi_{\a})+\l_{21}+N\eta)}\no\\
 &&\times \sum_{k=1}^N \a_{k}\lt\{\frac{\s(\eta)}
    {\s(\l_{21}+N\eta-\g_k)\s(\l_{21}+\eta+\g_k)}\rt\}^{N-1}
    {\rm det}{\cal{M}}(\g_k),\label{Det-1}
\eea where ${\cal{M}}(\g)$ is $N\times N$ matrix with matrix
elements are given by \bea
 {\cal{M}}(\g)_{\a\,j}=\frac{\s(u_{\a}-\xi_j+\l_{21}+\eta+\g)
  \s(u_{\a}-\xi_j+\l_{21}+N\eta-\g)}{\s(u_{\a}-\xi_j+\eta)\s(u_{\a}-\xi_j)},
  \quad \a,j=1,\ldots,N.\label{Det-2}
\eea The $2N$ parameters $\{\a_k;\g_k|k=1,\ldots,N\}$ in
(\ref{Det-1}) satisfy the following equations \bea
 &&\sum_{i=1}^N\a_i\lt\{\frac{\s(\l_{21}+\g_i)\s(\l_{21}-\g_i+(N-1)\eta)}
         {\s(\l_{21}+\g_i+\eta)\s(\l_{21}-\g_i+N\eta)}\rt\}^k
         \s(\l_{21}+\g_i+\eta)\s(\l_{21}-\g_i+N\eta)\no\\
 &&\qquad\qquad=
         \frac{\s(\l_{21})\s(\eta)\s(\l_{21}+(N-k)\eta)}
         {\s(\l_{21}+k\eta)},\quad
         k=0,\ldots N-1,\label{Restr-1}\\
&&\sum_{i=1}^N\a_i\lt\{\frac{\s(\l_{21}+\g_i)\s(\l_{21}-\g_i+(N-1)\eta)}
         {\s(\l_{21}+\g_i+\eta)\s(\l_{21}-\g_i+N\eta)}\rt\}^k
         \s(\g_i+(k+1-N)\eta)\s(k\eta-\g_i)\no\\
&&\qquad\qquad=0,\quad\quad
         k=0,\ldots N-1.\label{Restr-2}
\eea The constraints (\ref{Restr-1}) and (\ref{Restr-2}) assure
that the R.H.S. of (\ref{Det-1}) has the properties
(\ref{Property-1}),(\ref{Symmetry}), (\ref{quasi-per-1}),
(\ref{quasi-per-2}) and (\ref{Analytic}) of the partition function
$Z_N(\{u_{\a}\};\{\xi_j\};\l)$. Moreover,  one can check that the
R.H.S. of (\ref{Det-1}) also satisfies the relation
(\ref{property-2}). As is shown in section 5, these properties
uniquely determine the partition function. Therefore,
(\ref{Det-1}) gives an equivalent expression of  the partition
function (\ref{partition-function}).

\section*{Acknowledgments}
Financial support from the Australian Research Council is
gratefully acknowledged.

\vspace{1.00truecm}

\noindent{{\large \it Note added}}: \, After submitting our paper
to the arXiv, we became aware that the partition function for the
elliptic SOS model with the domain wall boundary condition has
been obtained in \cite{Ros08,Pak08,Raz08} by different methods.
Our approach is based   entirely on the algebraic Bethe ansatz
framework and can be used to obtain partition functions of the
eight-vertex model with open boundary conditions \cite{Yan09}. We
would like to thank V. Mangazeev for kindly drawing our attention
to these references.


\section*{Appendix A: The proof of (\ref{Symmetry})}
\setcounter{equation}{0}
\renewcommand{\theequation}{A.\arabic{equation}}
In this appendix, we show that $Z_N(\{u_{\a}\};\{\xi_i\};\l)$ is a
symmetric function of $\{u_{\a}\}$ and $\{\xi_i\}$ separatively.

Regarding  the tensor space indexed by $\bar{1},\ldots,\bar{N}$ as
the auxiliary space, (\ref{PF-1}) can be rewritten as \bea
 Z_N(\{u_{\a}\};\{\xi_i\};\l)
 &=&\langle\Omega^{(2)}(\l
    \hspace{-0.1truecm}+\hspace{-0.1truecm}\eta
    N\hat{2})|\,\tilde{\phi}^{\bar{1}}_{\l,\l-\eta\hat{1}}(u_1)
    T_{\bar{1}}(u_1)\phi^{\bar{1}}_{\l+\eta
    N\hat{2},\l+\eta(N-1)\hat{2}}(u_1)
    \ldots\no\\
 &&\times \tilde{\phi}^{\bar{N}}
    _{\l-\eta(N-1)\hat{1},\l-\eta N\hat{1}}(u_N)
    T_{\bar{N}}(u_N)\phi^{\bar{N}}_{\l+\eta\hat{2},\l}(u_N)
    |\Omega^{(1)}(\l)\rangle.\label{Proof-1}
\eea Following \cite{Yan04,Yan041}, let us introduce the face-type
monodromy matrix with elements given by \bea
 T(m;m_0|u)^j_{\mu}=\tilde{\phi}_{m+\eta\hat{\jmath},m}(u)\,T(u)\,
  \phi_{m_0+\eta\hat{\mu},m_0}(u),\quad j,\mu=1,2.\label{Proof-2}
\eea Then the partition function $Z_N(\{u_{\a}\};\{\xi_i\};\l)$
can be expressed in terms of the product of the matrix elements of
the face-type monodromy matrix
\bea
Z_N(\{u_{\a}\};\{\xi_i\};\l)\hspace{-0.26truecm}&=&\hspace{-0.26truecm}
    \langle\Omega^{(2)}(\l\hspace{-0.1truecm}+\hspace{-0.1truecm}\eta
    N\hat{2})|\,T(\l\hspace{-0.1truecm}-\hspace{-0.1truecm}\eta\hat{1};
    \l\hspace{-0.1truecm}+\hspace{-0.1truecm}\eta(N\hspace{-0.1truecm}-\hspace{-0.1truecm}1)
    \hat{2}|u_1)^1_2
    T(\l\hspace{-0.1truecm}-\hspace{-0.1truecm}2\eta\hat{1};
    \l\hspace{-0.1truecm}+\hspace{-0.1truecm}\eta(N\hspace{-0.1truecm}-\hspace{-0.1truecm}2)
    \hat{2}|u_2)^1_2\no\\
  &&\times
    \ldots T(\l-\eta N\hat{1};\l|u_N)^1_2\,
    |\Omega^{(1)}(\l)\rangle.\label{Proof-3}
\eea

The exchange relation (\ref{RLL-1}) and the vertex-face
correspondence relations (\ref{Face-vertex-1}) and
(\ref{Face-vertex-3}) enable us to derive the following exchange
relations for  the operators (\ref{Proof-2}) \bea
 &&\sum_{i,j=1}^2\R(u_1-u_2;m)_{ij}^{kl}
    T(m+\eta\hat{\jmath};m_0+\eta\hat{\nu}|u_1)^i_{\mu}
    T(m;m_0|u_2)^j_{\nu}\no\\
 &&\qquad\qquad =\sum_{\a,\b=1}^2\R(u_1-u_2;m_0)^{\a\b}_{\mu\nu}
    T(m+\eta\hat{k};m_0+\eta\hat{\a}|u_2)^l_{\b}
    T(m;m_0|u_1)^k_{\a}.\no
\eea For the case of $k=l=1$ and $\mu=\nu=2$, the above relations
become
\bea
 T(m+\eta\hat{1};m_0+\eta\hat{2}|u_1)^1_2\,T(m;m_0|u_2)^1_2=
   T(m+\eta\hat{1};m_0+\eta\hat{2}|u_2)^1_2\,T(m;m_0|u_1)^1_2.\label{Proof-4}
\eea This relation with  $T(m;m_0|u)^1_2$ given by (\ref{Proof-2})
implies that the partition function $Z_N(\{u_{\a}\};\{\xi_i\};\l)$
(\ref{Proof-3}) is a symmetric function of $\{u_{\a}\}$.
Similarly, one can check that the partition function is also a
symmetric function of $\{\xi_i\}$.


\section*{Appendix B: The proofs of (\ref{Residue}) and (\ref{property-2})}
\setcounter{equation}{0}
\renewcommand{\theequation}{B.\arabic{equation}}

Firstly, let us prove the analytic property (\ref{Residue}) of the
partition function. Since the partition function is a symmetric
function of $\{u_{\a}\}$ and $\{\xi_i\}$, it is sufficient to
prove (\ref{Residue}) for the case of $u_N=\xi_N-\eta$. For this
purpose, we need to rearrange the order of the product of
R-matrices in the expression (\ref{PF}) of the partition function
as follows
\bea
 \hspace{-1truecm}Z_N(\{u_{\a}\};\{\xi_i\};\l)
  \hspace{-0.32truecm}&=&\hspace{-0.32truecm}\langle\Omega^{(2)}(\l
    \hspace{-0.1truecm}+\hspace{-0.1truecm}\eta
    N\hat{2})|\langle\bar{\Omega}^{(1)}(\l)|\,
    R_{\bar{1},N}(u_1\hspace{-0.1truecm}-\hspace{-0.1truecm}\xi_N)
    R_{\bar{2},N}(u_2\hspace{-0.1truecm}-\hspace{-0.1truecm}\xi_N)
    \ldots R_{\bar{N},N}(u_N\hspace{-0.1truecm}-\hspace{-0.1truecm}\xi_N)
    \ldots\no\\
 &&\times R_{\bar{1},1}(u_1-\xi_1)\ldots
    R_{\bar{N},1}(u_N-\xi_1)|\bar{\Omega}^{(2)}(\l+\eta N\hat{2})\rangle
    |\Omega^{(1)}(\l)\rangle.\no\\
 &=&\hspace{-0.32truecm}\langle\Omega^{(2)}(\l
    \hspace{-0.1truecm}+\hspace{-0.1truecm}\eta
    N\hat{2})|\,\tilde{\phi}^{\bar{1}}_{\l,\l-\eta\hat{1}}(u_1)\ldots
    \tilde{\phi}^{\overline{N-1}}_{\l-\eta(N-2)\hat{1},\l-\eta(N-1)\hat{1}}(u_{N-1})\no\\
 &&\times R_{\bar{1},N}(u_1-\xi_N)\ldots
    R_{\overline{N-1},N}(u_{N-1}-\xi_N)\no\\
 && \times\tilde{\phi}^{\bar{N}}_{\l-\eta(N-1)\hat{1},\l-\eta
    N\hat{1}}(u_N)R_{\bar{N},N}(u_N-\xi_N)\phi^{N}_{\l-\eta(N-1)\hat{1},\l-\eta
    N\hat{1}}(\xi_N)\no\\
 &&\times R_{\bar{1},N-1}(u_1-\xi_{N-1})\ldots R_{\bar{N},N-1}(u_N-\xi_{N-1})
    \ldots R_{\bar{N},1}(u_N-\xi_{1})\no\\
 &&\times\phi^1_{\l,\l-\eta\hat{1}}(\xi_1)\ldots
    \phi^{N-1}_{\l-\eta(N-2)\hat{1},\l-\eta(N-1)\hat{1}}(\xi_{N-1})
    \,|\bar{\Omega}^{(2)}(\l+\eta N\hat{2})\rangle.\label{B-1}
\eea The expressions (\ref{Elements1})-(\ref{Elements2}) of the
matrix elements of $\R$ imply that \bea
 {\rm Res}_{u=-\eta}R(u;m)^{11}_{11}=0,\quad {\rm
  Res}_{u=-\eta}R(u;m)^{12}_{21}=
  \frac{\s(\eta)\s(m_{21}-\eta)}{\s'(0)\s(m_{21})}.
\eea Keeping the above equations in mind and using the vertex-face
correspondence relation (\ref{Face-vertex-2}), we find
\bea
 {\rm Res}_{u_N=\xi_N-\eta}
  \hspace{-0.36truecm}&&\hspace{-0.36truecm}Z_N(\{u_{\a}\};\{\xi_i\};\l)\no\\
 &=&
  \frac{\s(\eta)\s(\l_{21}+(N-1)\eta)}{\s'(0)\s(\l_{21}+N\eta)}
  \langle\Omega^{(2)}(\l
    \hspace{-0.1truecm}+\hspace{-0.1truecm}\eta
    N\hat{2})|\no\\
  &&\times\tilde{\phi}^{\bar{1}}_{\l,\l-\eta\hat{1}}(u_1)\ldots
    \tilde{\phi}^{\overline{N-2}}_{\l-\eta(N-3)\hat{1},\l-\eta(N-2)\hat{1}}(u_{N-2})\,\,
    \tilde{\phi}^{\overline{N}}_{\l+\eta\hat{2}-\eta(N-1)\hat{1},\l-\eta(N-1)\hat{1}}(u_{N})
    \no\\
  &&\times R_{\bar{1},N}(u_1-\xi_N)\ldots
    R_{\overline{N-2},N}(u_{N-2}-\xi_N)\no\\
  &&\times \tilde{\phi}^{\overline{N-1}}
    _{\l-\eta(N-2)\hat{1},\l-\eta(N-1)\hat{1}}(u_{N-1})
    R_{\overline{N-1},N}(u_{N-1}\hspace{-0.1truecm}-\hspace{-0.1truecm}\xi_N)
    \phi^N_{\l+\eta\hat{2}-\eta(N-1)\hat{1},\l-\eta(N-1)\hat{1}}(\xi_N)\no\\
  &&\times R_{\bar{1},N-1}(u_1-\xi_{N-1})\ldots R_{\bar{N},N-1}(u_N-\xi_{N-1})
    \ldots R_{\bar{N},1}(u_N-\xi_{1})\no\\
  &&\times\phi^1_{\l,\l-\eta\hat{1}}(\xi_1)\ldots
    \phi^{N-1}_{\l-\eta(N-2)\hat{1},\l-\eta(N-1)\hat{1}}(\xi_{N-1})
    \,|\bar{\Omega}^{(2)}(\l+\eta N\hat{2})\rangle\no\\
 &=&\no\\
 &\vdots&\no\\
 &=&\hspace{-0.32truecm}
  \frac{\s(\eta)\s(\l_{21}+(N-1)\eta)}{\s'(0)\s(\l_{21}+N\eta)}
  \prod_{l=1}^{N-1}\R(u_l-\xi_N;\l-\eta l\hat{1})^{12}_{12}\no\\
  &&\times \langle\Omega^{(2)}(\l
    \hspace{-0.1truecm}+\hspace{-0.1truecm}\eta
    N\hat{2})|\,\phi^N_{\l+\eta\hat{2},\l}(\xi_N)\no\\
  &&\times\tilde{\phi}^{\bar{1}}_{\l+\eta\hat{2},\l+\eta\hat{2}-\eta\hat{1}}(u_1)\ldots
    \tilde{\phi}^{\overline{N-1}}_{\l+\eta\hat{2}-\eta(N-2)\hat{1},
    \l+\eta\hat{2}-\eta(N-1)\hat{1}}(u_{N-1})\no\\
  &&\times R_{\bar{1},N-1}(u_1-\xi_{N-1})\ldots
    R_{\overline{N-1},N-1}(u_{N-1}-\xi_{N-1})\no\\
  &&\times \tilde{\phi}^{\bar{N}}_{\l+\eta\hat{2}-\eta(N-1)\hat{1},
    \l-\eta(N-1)\hat{1}}(u_{N})
    R_{\bar{N},N-1}(u_N\hspace{-0.1truecm}-\hspace{-0.1truecm}\xi_{N-1})
    \phi^{N-1}_{\l-\eta(N-2)\hat{1},\l-\eta(N-1)\hat{1}}(\xi_{N-1})\no\\
  &&\times R_{\bar{1},N-2}(u_1-\xi_{N-2})\ldots
    R_{\bar{N},N-2}(u_{N}-\xi_{N-2})\ldots
    R_{\bar{N},1}(u_N-\xi_1)\no\\
  &&\times\phi^1_{\l,\l-\eta\hat{1}}(\xi_1)\ldots
    \phi^{N-2}_{\l-\eta(N-3)\hat{1},\l-\eta(N-2)\hat{1}}(\xi_{N-2})
    \,|\bar{\Omega}^{(2)}(\l+\eta N\hat{2})\rangle\no\\
 &=&\no\\
 &\vdots&\no\\
 &=&\hspace{-0.32truecm}
  \frac{\s(\eta)\s(\l_{21}+(N-1)\eta)}{\s'(0)\s(\l_{21}+N\eta)}
  \prod_{l=1}^{N-1}\R(u_l-\xi_N;\l-\eta l\hat{1})^{12}_{12}
  \R(u_N-\xi_l;\l-\eta l\hat{1})^{21}_{21}\no\\
  &&\times \langle\Omega^{(2)}(\l
    \hspace{-0.1truecm}+\hspace{-0.1truecm}\eta
    N\hat{2})|\,\phi^N_{\l+\eta\hat{2},\l}(\xi_N)\no\\
  &&\times\tilde{\phi}^{\bar{1}}_{\l+\eta\hat{2},\l+\eta\hat{2}-\eta\hat{1}}(u_1)\ldots
    \tilde{\phi}^{\overline{N-1}}_{\l+\eta\hat{2}-\eta(N-2)\hat{1},
    \l+\eta\hat{2}-\eta(N-1)\hat{1}}(u_{N-1})\no\\
  &&\times R_{\bar{1},N-1}(u_1-\xi_{N-1})\ldots
    R_{\overline{N-1},N-1}(u_{N-1}-\xi_{N-1})\ldots\no\\
  &&\times R_{\bar{1},1}(u_1-\xi_{1})\ldots
    R_{\overline{N-1},1}(u_{N-1}-\xi_{1})\no\\
  &&\times\phi^1_{\l+\eta\hat{2},\l+\eta\hat{2}-\eta\hat{1}}(\xi_1)\ldots
    \phi^{N-1}_{\l+\eta\hat{2}-\eta(N-2)\hat{1},\l+\eta\hat{2}-\eta(N-1)\hat{1}}(\xi_{N-1})\no\\
  &&\times\tilde{\phi}^{\bar{N}}_{\l+\eta\hat{2},\l}(u_N)\,
     |\bar{\Omega}^{(2)}(\l+\eta N\hat{2})\rangle.
\eea It is understood that $u_N=\xi_N-\eta$ in the above
equations. With help of the definitions (\ref{Boundary-state-2})
and (\ref{Boundary-state-3}) of the boundary states and the
condition (\ref{Int2}), we finally obtain the residue of
$Z_N(\{u_{\a}\};\{\xi_i\};\l)$ at the simple pole $\xi_N-\eta$:
\bea
 {\rm Res}_{u_N=\xi_N-\eta}
  \hspace{-0.36truecm}&&\hspace{-0.36truecm}Z_N(\{u_{\a}\};\{\xi_i\};\l)\no\\
   &=&\hspace{-0.32truecm}
      \frac{\s(\eta)\s(\l_{21})}{\s'(0)\s(\l_{21}+\eta)}
      \prod_{l=1}^{N-1}\frac{\s(u_l-\xi_N)}{\s(u_l-\xi_N+\eta)}
      \prod_{j=1}^{N-1}\frac{\s(\xi_j-\xi_N+\eta)}{\s(\xi_j-\xi_N)}\no\\
   &&\times\tilde{\phi}^1_{\l+\eta\hat{2}+\eta(N-1)\hat{2},
      \l+\eta\hat{2}+\eta(N-2)\hat{2}}(\xi_1)\ldots
      \tilde{\phi}^{N-1}_{\l+\eta\hat{2}+\eta\hat{2},\l+\eta\hat{2}}(\xi_{N-1})\no\\
   &&\times\tilde{\phi}^{\bar{1}}_{\l+\eta\hat{2},\l+\eta\hat{2}-\eta\hat{1}}(u_1)\ldots
    \tilde{\phi}^{\overline{N-1}}_{\l+\eta\hat{2}-\eta(N-2)\hat{1},
    \l+\eta\hat{2}-\eta(N-1)\hat{1}}(u_{N-1})\no\\
  &&\times R_{\bar{1},N-1}(u_1-\xi_{N-1})\ldots
    R_{\overline{N-1},N-1}(u_{N-1}-\xi_{N-1})\ldots\no\\
  &&\times R_{\bar{1},1}(u_1-\xi_{1})\ldots
    R_{\overline{N-1},1}(u_{N-1}-\xi_{1})\no\\
  &&\times\phi^{\bar{1}}_{\l+\eta\hat{2}+\eta(N-1)\hat{2},
  \l+\eta\hat{2}+\eta(N-2)\hat{2}}(u_1)\ldots
    \phi^{\overline{N-1}}_{\l+\eta\hat{2}+\eta\hat{2},\l+\eta\hat{2}}(u_{N-1})\no\\
  &&\times\phi^{1}_{\l+\eta\hat{2},\l+\eta\hat{2}-\eta\hat{1}}(\xi_1)\ldots
    \phi^{N-1}_{\l+\eta\hat{2}-\eta(N-2)\hat{1},\l+\eta\hat{2}-\eta(N-1)\hat{1}}(\xi_{N-1})\no\\
  &=&\hspace{-0.32truecm}
      \frac{\s(\eta)\s(\l_{21})}{\s'(0)\s(\l_{21}+\eta)}
      \prod_{l=1}^{N-1}\frac{\s(u_l-\xi_N)}{\s(u_l-\xi_N+\eta)}
      \prod_{j=1}^{N-1}\frac{\s(\xi_j-\xi_N+\eta)}{\s(\xi_j-\xi_N)}\no\\
  &&\times Z_{N-1}(\{u_{\a}\}_{\a\neq N};\{\xi_i\}_{i\neq
      N};\l+\eta\hat{2}).\label{B-4}
 \eea Therefore, we have completed the proof of (\ref{Residue}).

Noting $\R(u;m)^{ii}_{ii}=1$ for any values of $u$, $m$ and
$i=1,2$ and  the initial condition (\ref{Initial}) of $R$, by a
similar procedure as above, one can show \footnote{In the proof of
(\ref{B-5}), it is convenient to keep the same order of the
product of the R-matrices as that of (\ref{PF}) (c.f. (\ref{B-1}))
in the calculation.} that \bea
 Z_N(\{u_{\a}\};\{\xi_i\};\l)\lt|_{u_N=\xi_1}\rt.=
  Z_{N-1}(\{u_{\a}\}_{\a\neq N};\{\xi_i\}_{i\neq
  1};\l).\label{B-5}
\eea The fact that $Z_N(\{u_{\a}\};\{\xi_i\};\l)$ is a symmetric
function of $\{u_{\a}\}$ and $\{\xi_i\}$ then leads to
(\ref{property-2}).


\end{document}